\documentclass[twocolumn,preprintnumbers,amsmath,amssymb]{revtex4}

\usepackage{graphicx}
\usepackage{dcolumn}
\usepackage{bm}
\newcommand \be  {\begin{equation}}
\newcommand \bea {\begin{eqnarray} \nonumber }
\newcommand \ee  {\end{equation}}
\newcommand \eea {\end{eqnarray}}

\begin{document}


\title{On the Adam-Gibbs-Wolynes scenario for the viscosity increase in
glasses}

\author{Jean-Philippe Bouchaud}
\email{bouchau@spec.saclay.cea.fr}
\author{Giulio Biroli}%
 \email{biroli@spht.saclay.cea.fr}
 \affiliation{$^{*}$Service de Physique de l'{\'E}tat Condens{\'e}, Orme
 des Merisiers, 
 CEA Saclay, 91191 Gif sur Yvette Cedex, France.\\
 $^{*}$ Science \& Finance, Capital Fund Management, 6-8 Bd Haussmann,
 75009 Paris, France.\\
 $^{\dagger}$Service de Physique Th{\'e}orique, Orme des Merisiers, CEA
 Saclay, 
 91191 Gif sur Yvette Cedex, France.
 }%

 \date{\today}

 \begin{abstract}
 We reformulate the interpretation 
 of the mean-field glass transition scenario for finite
 dimensional systems, proposed by Wolynes and collaborators. 
 This allows us to establish clearly a temperature dependent length
 $\xi^{*}$ above which the mean-field glass
 transition picture has to be modified. We argue in favor of the mosaic state 
 introduced by Wolynes and collaborators, which leads to the Adam-Gibbs 
 relation between the viscosity and configurational 
 entropy of glass forming liquids. 
 Our argument is a mixture of 
 thermodynamics and kinetics, partly inspired by the Random Energy
 Model: small clusters of particles 
 are thermodynamically frozen in low energy states, 
 whereas large clusters are kinetically frozen by
 large activation energies. The relevant relaxation time is that of 
 the smallest `liquid' clusters. Some physical consequences are discussed.
 \end{abstract}

 \pacs{Valid PACS appear here}

\maketitle
\section{Introduction} 

One of the most striking properties of `fragile' glasses is the
extremely fast rise of their viscosity $\eta$ 
(or their structural relaxation time $\tau$),
that increases by 10 orders of magnitude as the temperature is decreased
by less than a factor 2 \cite{Nature}. The 
increase is faster than thermal Arrhenius activation 
over a constant energy barrier, and can be well-fit by the so-called
Vogel-Fulcher (VF) equation \cite{VF}:
\be
\log \eta = \log \eta_0 + \frac{\Delta}{T-T_{VF}},
\ee
at least in the range where the system can be equilibrated and the
viscosity measured. Clearly, this becomes 
impossible when $T$ approaches $T_{VF}$, and the above fit cannot be
tested in the immediate vicinity of the
predicted divergence. Correspondingly, other functional forms cannot be
ruled out, such as a generalized VF 
form:
\be\label{VFG}
\log \eta = \log \eta_0 +
\left(\frac{\Delta}{T-T_{\gamma}}\right)^\gamma.
\ee
For example, many experimental results can also be reproduced with
$\gamma=2$, $T_\gamma=0$ \cite{Bassler}. 
Many other functional
forms have been proposed, see for example \cite{Tarjus,BG}, with
equivalent goodness-of-fits. However, a 
rather remarkable aspect of the original Vogel-Fulcher equation is that
the extrapolated freezing temperature $T_{VF}$ is
found to be very close, for a whole range of materials, to the Kauzmann
temperature $T_K$ where the extrapolated 
entropy of the glass becomes
smaller than that of the crystal. More precisely, the ratio $T_K/T_{VF}$
is
in the range $0.9 - 1.1$ for a whole slew of glass formers, with $T_K$
itself changing from $50$ K to $1000$ K --
see \cite{Angell}. This ``coincidence'' suggests that there might be
some
truth in the Vogel-Fulcher fit as well as in the extrapolation leading
to $T_{K}$. It forcefully points towards an explanation of the
viscosity increase in terms of a thermodynamic critical point, albeit
of a non conventional type (for an interesting discussion of this 
point of view, see e.g. \cite{Sethna}). This perspective has attracted a
great deal of 
attention over many years, 
starting from the early work
of Adam, Gibbs and Di Marzio \cite{GdM,AG,Johari}. In particular, 
the theory of Adam and Gibbs predicts a relation between
viscosity and configurational entropy of the glass $s_c(T)$ given by:
\be\label{AG}
\log \eta = \log \eta_0 + \frac{\Delta}{Ts_c(T)}.
\ee
This relation is in rather good quantitative agreement with many
experimental results \cite{Angell,Hodge,Johari,new}. 
Qualitatively, the argument is that as the configurational entropy of
the glass goes to zero, there are less and less available 
configurations to move the molecules around, and the dynamics slows
down. However, the Adam-Gibbs theoretical argument
is far from being water-tight, let alone convincing (see below). A
related, but 
distinct, argument was proposed 
in the late 80's by Kirkpatrick, Thirumalai and Wolynes
\cite{Wolynes}, and repeated (and partially reformulated) by Wolynes
and collaborators in different contexts since then \cite{W2,W3,W4}. 
The basic ingredient of this theory is the nucleation of so-called 
`entropic droplets' between different
metastable configurations of the super-cooled liquid. Despite several
quantitative successes reported in the literature
\cite{W3,W4}, it is fair to say that the entropic droplet scenario is
still to convince many workers in the field. 
One of the reasons is that the physics behind `the entropic driving
force' leading to nucleation is rather obscure.
The aim of this note is to propose a clearer 
-- at least to our eyes -- and somewhat different interpretation of the
scenario of Wolynes et al. that leads to the Adam-Gibbs relation. 
A possible mechanism underlying the dramatic slow down of 
super-cooled liquids and the physical interpretation of the glass state
appears clearly. Some semi-quantitative predictions
of this picture are spelled out, and could be checked in numerical 
simulations and experiments.  

\section{The Adam-Gibbs and Wolynes arguments revisited} 

Adam and Gibbs envision the super-cooled liquid as progressively
organizing in larger and larger cooperative regions 
that have to collectively rearrange. Each of these cooperative regions,
of linear size $\xi$, only has a few number 
$\Omega$
of preferred configurations, where $\Omega$ is independent of $\xi$ --
say $\Omega=2$ (on this point, see \cite{Johari}). 
The total entropy of the 
super-cooled liquid is therefore, for a volume $V$:
\be
V s_c(T) = k_{B}\left(\frac{V}{\xi^3}\right) a^3 \log \Omega,
\ee
where $a^3$ is the volume of a single molecule that we will set to $a=1$
in the following. 
The next step is to {\it assume} that the energy barrier $B$ for 
rearranging a region of size $\xi$ scales with the total number of
molecules in that region:
\be
B = \Delta_0 {\xi}^3.
\ee
Therefore, the time $\tau$ to collectively rearrange that region is
given by:
\be
\log \tau = \log \tau_0 + \frac{\Delta_0 \log \Omega}{Ts_c(T)}.
\ee
Identifying the viscosity and the relaxation time leads to the
Adam-Gibbs relation Eq. (\ref{AG}) with $\Delta
= \Delta_0 \log \Omega$. The difficulties in the above argument are 
that (a) it seems very unnatural to assume that
the number of metastable configurations in a volume $\xi^3$ is
independent of $\xi$. One would rather (see below) 
more naturally expect that this number is in fact exponential in
$\xi^3$. (b) the barrier for rearranging a region
of size $\xi$ should scale as $\xi^\psi$ with $\psi \leq d = 3$. The
limit $\psi=d$ implies that the only possible relaxation mechanism is
a cooperative movement involving {\it a finite fraction of all the
particles}, which is unlikely for large $\xi$.
In spite of these deficiencies, the Adam-Gibbs relation
fares quite well in accounting for experimental data
\cite{Angell,Hodge,Johari,new}, 
and the original Adam-Gibbs work has had (and still has) an enormous impact.

Another line of thought was initiated in a series of remarkable papers
by Kirkpatrick, Thirumalai and Wolynes \cite{Wolynes}, 
that established
a profound analogy between super-cooled liquids and the physics of a
family of mean-field spin-glasses 
\cite{Review}, which exhibit two
transition temperatures. One is a dynamical transition temperature where
(in mean-field) ergodicity is broken and
the system is trapped close to metastable states of free-energy larger
than that of the paramagnetic (liquid) state, 
which is still the thermodynamically stable state. This
transition temperature can formally be identified with the Mode-Coupling
transition (MCT) temperature $T_{MCT}$
\cite{Gotze,Wolynes}. The second transition,
at a lower temperature $T_0$, corresponds to a {\it bona-fide} 
thermodynamical phase transition to a glassy phase,
with no latent heat but a discontinuous order parameter (a transition
called `discontinuous' or `random first order').
Between $T_0$ and $T_{MCT}$, the difference in free-energy between
individual metastable states and the liquid state 
is found to be exactly the `complexity', 
i.e. the log-degeneracy of the metastable states, that one can therefore
associate to the configurational entropy of the liquid
\footnote{Depending on 
the author, the meaning of
configurational entropy appears to be different. Here and in the
following
we call configurational entropy the logarithm of the number of
metastable states. See \cite{Wolynes,BiroliMonasson} for
discussions.}. Roughly speaking, the total entropy of the liquid is 
made of a `vibrational',
bottom of valley contribution (that also exists, and is of similar
magnitude, in the frozen glass or the solid phase),
plus this configurational entropy that reflects the evolution of the
liquid from one quasi-frozen configuration to
another -- an evolution that becomes impossible, in mean-field, when $T
\leq T_{MCT}$. Since the configurational entropy 
vanishes continuously at $T_0$ in these mean-field models, it is
tempting to associate $T_0$ with the Kauzmann temperature 
$T_K$. 
In finite dimensions, the mode coupling transition becomes a
cross-over since barriers between metastable states
are finite and some `activated' dynamics still occurs below $T_{MCT}$. 
Furthermore, the thermodynamic ideal glass transition at $T_{K}$ is 
avoided in many real cases (because of crystal nucleation for
instance) but this turns out to be irrelevant for practical purposes \cite{f1}
(see also the discussion in \cite{Wolynes}).
Wolynes and
collaborators argue that
the transitions between different states occur via nucleation
\cite{Wolynes}, and that this process is responsible for the dramatic
increase of the
structural relaxation timescale. However, contrarily to usual nucleation
where the driving force comes from a difference of bulk free-energy
between the typical metastable state sampled at equilibrium and the
invading phase, 
all metastable states here have the same bulk free-energy. The driving
force is argued to be `entropic', and given by
the log-degeneracy of all {\it possible} phases; for a droplet of size
$\xi$, the entropy gain would then be (in
$d$ dimensions)
$-T s_c(T) \xi^d$, whereas the energy loss due to a mismatch between the
nucleating phase and the surrounding state 
is given by a generalized surface tension $\Upsilon \xi^\theta$, with
$\theta \leq d-1$. The two contributions 
balance when $\xi^{*(d-\theta)}=\Upsilon/T s_c(T)$. 
This length is interpreted \cite{Wolynes,W2,W3} as the
typical size of a 
`mosaic' state, that pictures the super-cooled liquid as a patchwork of
local metastable 
configurations. The free-energy barrier coming from the balance between
the entropic driving force and the
surface tension leads, through the Arrhenius relation, to a generalized
Adam-Gibbs relation (in $d$ dimensions):
\be\label{AGW}
\log \tau = \log \tau_0 + c \frac{\Upsilon}{k_B T}
\left(\frac{\Upsilon}
{Ts_c(T)}\right)^{\frac{\theta}{d-\theta}},
\ee
where $c$ is a model dependent constant. 
Using $s_c \sim T-T_K$, this corresponds to a generalized
Vogel-Fulcher law, Eq. (\ref{VFG}), with
$\gamma=\theta/(d-\theta)$. It was argued in \cite{Wolynes} that
$\theta=3/2$ in three dimension and as a consequence one gets back to
the usual Vogel-Fulcher and Adam-Gibbs laws.
The obscure point of the above argument is the precise nature of the
`entropic driving force', in particular the way 
it mixes in a subtle manner static (thermodynamical) and dynamical
considerations. Indeed, 
assuming the different cells of the
mosaic state to be independent, the total entropy of the system is
actually:
\be
{\cal S} = \frac{V}{\xi^d} s_c(T) \xi^d,
\ee
which is independent of $\xi$, precisely because the entropy of each
cell scales like $\xi^d$ (at variance with
the Adam-Gibbs picture, where it is independent of $\xi$). 
So, why should the system break up in different domains and
pay surface tension? What exactly fixes the scale over which 
cooperative events take place? What is the physical meaning of 
the entropic driving force?

We believe that the picture of Wolynes et al. is in fact fundamentally
correct, although the arguments supporting it are
somewhat unsatisfactory. The aim of the following sections is to
establish 
what we hope to be a 
more convincing version of these arguments. The following considerations
actually share some similarities, but also
important differences, with a very recent preprint by Lubchenko and
Wolynes \cite{W6}. 

\section{What is a glass ? A view from the Random Energy Model}

\subsection{The glass transition in the Random Energy Model}

A true glass has a solid-like response to shear and has thus an infinite
viscosity. Since the viscosity is the integral
over time of the stress correlation function, a glass is characterized
by the fact that local stress 
fluctuations do not decorrelate in time. This is possible if the
particles only visit, with appreciable probability,
a finite number of configurations, a situation characteristic of {\it
broken ergodicity} \cite{GG}. If all configurations 
can 
be probed with appreciable probabilities, then all correlation functions
tend to zero in the long time limit: 
ergodicity is restored. 

A simple, but illuminating model for such a behaviour is the Random
Energy Model (REM), where the $2^N$ 
configurations have random, independent energies \cite{REM,BM}. 
For $T > T_K$, the entropy is non zero, which means that the
system explores an exponential number (in $N$) of different states, each
of which during a vanishing fraction of the
total time. For $T < T_K$, on the other hand, the entropy is zero, and
the system is localized in a finite number 
of configurations. More precisely, there are a finite number of
configurations in which the system spends a 
finite fraction of the total time $t_w$, however long $t_w$ might be
\cite{B92,MB,BenA}. 
At low temperatures, the energy dominates -- even 
if there are $2^N$ states that await the system, only the very low
states have an appreciable probability to be
occupied. As soon as $T$ crosses the value $T_K$, however, the system is
`sucked up' by higher energy states because 
of their huge number. Because the number of visited states is finite for
$T < T_K$, the system cannot forget its past,
correlation functions do not decay to zero, one has a glass. For $T >
T_K$, on the other hand, the system can lose 
itself in a large number of states, such that the probability to come
back to an already visited state is close to
naught. Autocorrelations then go to zero, the system is liquid. 
We shall show in the following that a similar mechanism is behind the
existence
of the mosaic state and fixes the length $\xi$, which actually plays
the very same r{\^o}le as temperature in the Random Energy Model.

\subsection{Entropy driven cluster melting}

Consider a large ensemble of interacting particles that become glassy at
low temperatures, for example a binary
Lennard-Jones system or soft spheres, etc..
Let us assume that the liquid is trapped in one -- call it $\alpha $ --
of the (exponentially numerous) 
metastable states found in qualitative \cite{Wolynes} and quantitative 
\cite{W5,MP,Cardenas} mean-field-like studies. 
In the following we shall establish the length above
which this assumption is inconsistent, providing an upper bound for the
typical length-scale of the mosaic state. 
We freeze the motion of all particles outside a spherical region of
radius $\xi$ and focus on the thermodynamics of the small cluster of
particles inside the sphere, ${\cal C}(\xi)$, 
subject to the boundary conditions imposed by the frozen particles
outside 
the sphere. Because of the `pinning' field 
imposed by these frozen particles, some configurations of ${\cal
C}(\xi)$ are particularly favored energetically. 
When $\xi^{d} s_{c}$ is much larger than unity there are many
metastable states for the particles in the cluster. Their number is 
\be
{\cal N} \approx \exp[\xi^d s(f,T)/k_B]
\ee
where $f$ is the excess free energy per unit volume 
(counted from the ground state), and
$s$ the configurational entropy per unit volume at that free-energy. 
The boundary condition imposed by the external particles, frozen
in state $\alpha$, act as a random boundary 
field for all other metastable states
except $\alpha$ itself, for which these boundary conditions perfectly
match. As a consequence
the partition function of the cluster 
${\cal C}(\xi)$ is given by the sum of two contributions:
\begin{eqnarray}
Z(\xi,T)\approx
\sum_{\beta \neq \alpha }\exp [-\xi^{d} \frac{f_{\beta
}}{k_{B}T}]+\exp[-\xi^d\frac{
f_{\alpha }}{k_{B}T}+\frac{\Upsilon \xi^{\theta}}{k_B T}]\approx
\nonumber
\\ 
\int_0^\infty df\exp\left[\frac{\xi^d\{T s(f,T)-f\}}{k_B T}\right]+
\exp
\left[-\frac{\xi^d
f_{\alpha }}{k_{B}T}+\frac{\Upsilon \xi^{\theta}}{k_B
T}\right]\label{z2}
\end{eqnarray}
where $f_{\beta}$ is the excess free energy per unit volume 
that the state $\beta$ would
have with a typical, random boundary field created by another 
state, and the term $\Upsilon
\xi^{\theta}$ is the free energy gain due to the matching boundary
condition (we have dropped the irrelevant factor $\exp
(-\xi^{d}f_{0}/k_{B}T)$ where $f_{0}$ is the ground-state free energy).
We focus on the case $T$ close to $T_{K}$ and study the stability
against
`fragmentation' of
a typical state $\alpha$ at that temperature, i.e. a state with
excess free energy $f^{*}$ equal to the one dominating the integral
in (\ref{z2}), such that $ds/df^{*}=1/T$. (States with a higher
free-energy
have very small probabilities to be observed at that temperature, and
will
typically not appear in the mosaic state. We will comment below on
states
with free energy lower than $f^*$).
The partition function of the cluster immersed in the $\alpha$ 
state, $Z(\xi,T)$, then reads:
\be\label{final}
Z(\xi,T) \approx \exp[-\xi^{d} \frac{f^*}{k_{B}T}] 
\left(\exp [\xi^{d} \frac{s^*}{k_{B}}]
+ \exp [\frac{\Upsilon \xi^{\theta}}{k_B T}] \right).
\ee
The above expression is central to our argumentation.  
It is easy to see that when $\xi$ is small enough, the second term, 
which is the 
contribution of the `matched' state $\alpha$, dominates the partition 
function, as long as $\theta < d$. On the contrary, the first term 
becomes overwhelming for larger $\xi$, and gives back exactly the free
energy of the {\it liquid} state.  
Physically the system has the choice of either being in a 
single `matched' state,
losing configurational entropy but gaining the boundary free energy
term,
or being distributed over all the other
states, thereby gaining the configurational entropy but losing the 
boundary term.  In mean-field models, the 
entropy $s(f,T)$, as in the REM, vanishes linearly when $f \to 0$, with
a
negative curvature. This generically leads to an entropy $s^{*}$ (and a
free-energy $f^*$) that behave
as $T - T_K$, in the vicinity of $T_K$. 
The cross-over between the two regimes therefore takes 
place for $(\xi^{*})^{d-\theta} \propto \Upsilon/k_B(T-T_{K})$. 
Assuming that energy barriers grow like $\xi^\psi$, one
finds a typical equilibration time for the region ${\cal C}(\xi)$ given
by:
\be\label{tauxi}
\tau(\xi) \approx \tau_0 \exp\left(\frac{\Delta_0 \xi^\psi}{k_B
T}\right).
\ee
For a frozen environment, the situation is therefore as follows: for
small $\xi < \xi^*$, the dynamics from state to
state is fast (low barriers) but leads to nowhere -- the system ends up
always visiting the same state. For larger
$\xi > \xi^*$, the system can at last delocalize itself in phase space
and kill correlations, but this takes an 
increasingly large time. {\it Thermodynamically}, large clusters are in
the
liquid state: see Eq. (\ref{final}).

\subsection{Relaxation times in the mosaic state}

Since the hypothesis that the region ${\cal C}(\xi)$ is in the same
state as its environment necessarily breaks down at $\xi^{*}$, it is
reasonable to assume that this length-scale
should be identified as the typical length-scale of the mosaic state.
Note that a trivial lower bound for $\xi^{*}$ is 
given by $s_{c}(T)^{-1/d}$, which coincides with the Adam-Gibbs
prediction and comes from the fact that for smaller length-scales there
are typically no states available.\\

Thus, we came up with a scenario in which
relaxation modes of length $\xi < \xi^*$ cannot be used to restore 
ergodicity in the system. 
The configurational entropy on these scales is too small to stir the
configurations efficiently and win over the dynamically generated
pinning
field due to the environment. The motion on these 
length scales corresponds to a generalized `cage' effect and contributes
to the $\beta$-relaxation. Conversely, 
on scales $\xi > \xi^*$, the exploration of the different available
states in nearby regions leads to a self-generated pinning field 
that decorrelates to zero, after a time scale $\sim \tau(\xi)$,
i.e. ergodicity is restored by large scale modes $\xi > \xi^*$.
From the above arguments, it is thus clear that the relaxation time is
$\tau(\xi^*)$: smaller length scales are faster 
but unable to decorrelate, whereas larger scales are orders of magnitude
slower so that the evolution on these scales 
will be short-circuited by a relaxation in parallel of smaller blobs of
size $\xi^*$. Using $\xi^* \sim 
s_{c}^{1/(\theta-d)}$ and Eq. (\ref{tauxi}), we finally recover
a result similar to Eq. (\ref{AGW}) with the exponent
${\theta}/{(d-\theta)}$ replaced by ${\psi}/{(d-\theta)}$.\\

The values of $\theta$ and $\psi$ should of course be calculated in the
framework of a precise model. On general grounds one expects that they 
verify the inequalities $\theta \leq d-1$ and $\theta \leq \psi \leq
d-1$ \cite{FH}. If $\theta \leq 0$ the system is at or below its lower
critical dimension
and the mean-field picture is completely wiped out.
Since $\xi^*$
is not expected to be much larger than $\sim 6-20$ in physical
situations, 
these exponents are anyway not very accurately 
defined. Furthermore, as discussed in \cite{W4}, the surface tension
$\Upsilon$
is expected to vanish at $T_{MCT}$ since at that temperature the
metastable states become marginally stable. As
a consequence, $\Upsilon$ increases from zero to a finite value
as the temperature decreases from $T_{MCT}$ to $T_{K}$. This leads, for
the region
of experimental interest where $T$ is somewhat larger than $T_K$, to an
effective exponent $\gamma$ larger 
than the `true' one in the immediate vicinity of $T_{K}$ \cite{f1}.
Finally, the value of $\gamma=\psi/(d-\theta)$ can vary quite a 
bit away from its Vogel-Fulcher value $\gamma=1$ without changing
dramatically 
the `critical' temperature extrapolated from experimental data. For
example, 
using $\gamma=1/2$ would lead to a critical temperature only 
$5 - 10 \%$ above the 
Vogel-Fulcher temperature, and would not strongly affect the near coincidence
between $T_{\gamma}$ 
and $T_K$. \\

The above argument gives the typical relaxation time in the system,
relevant for example to
compute the diffusion constant $D$ of a tracer particle, expected to
behave as $D \sim \xi^{*2}/\tau(\xi^*)$.
On the other hand, the viscosity is proportional to the {\it average}
relaxation time, which may in fact 
be dominated by the slowest regions in the system. This is related to
the question of low energy states, with 
$0 \leq f < f^*$, that we did not consider above. Repeating the above
arguments, one finds that these states must 
fragment on a scale larger than $\xi^*$. (More precisely, if $f=u f^*$,
the scale $\xi_u$ is given 
by $\xi^*/(u+\varepsilon)^{1/(d-\theta)}$ with
$\varepsilon=(T-T_K)/T_K$.) 
The reason is quite simple: since they are (free-)energetically
favored one has to reach a much larger size to balance the 
boundary free-energy term and the entropy gained by distributing
the system over all the other states. 
What this implies is that there are, at
any instant of time, regions of 
space corresponding to these low free-energy states that will be
characterized by a somewhat
larger length scales, but considerably larger relaxation times. While
$\xi^*$ is the typical length-scale
of the mosaic state, we expect a distribution of length-scales up to 
a cut-off $\xi_{\max} \propto \xi^{*}/\varepsilon^{1/(d-\theta)}$, given
by the
fragmentation length of the lowest free-energy states \footnote{We
assumed that the configurational entropy vanishes linearly in $f$ and
that $\partial S/\partial T \leq 0$ at least close to the lowest 
free-energy states}.
Correspondingly, one has a very broad spectrum of relaxation times and
strong
dynamical heterogeneities: not surprisingly, the deepest states are
slowest to fragment 
and relax. Since the ratio $\xi_{\max}/\xi^*$ diverges as $T \to T_K$,
one finds that (a) the 
ratio of the average relaxation time $\langle \tau \rangle$ to the
typical relaxation time 
$\tau(\xi^*)$ diverges as $T \to T_K$, accounting for the observed
decoupling between viscosity 
and diffusion \cite{Decoupling} and (b) the width of the distribution of
the logarithm of relaxation times increases 
as the temperature is decreased, meaning that relaxation functions
become more and more stretched as $T \to T_K$, 
again a common experimental observation. Within this framework, we also
recover naturally the 
correlation between the stretching exponent $\beta$ and fragility
reported in \cite{W3}. \\

Concluding this section, we note that the mechanism proposed in this
section 
to relate the structural relaxation timescale to the behavior of the
configurational entropy is
different from the one of Adam-Gibbs. It is based, physically, on the
competition between configurational entropy and dynamically 
generated pinning field. The very same competition induces a phase
transition in the Random Energy Model.
Although our argumentation is somewhat different from that of \cite{W6},
it leads to very similar physical conclusions.

\section{Remarks on the length $\xi^*$ and generality of the
mosaic state}\label{mosaic}

The conclusion of the previous section as well as the results of
\cite{Wolynes,W2,W3,W4} indicate that a system characterized by a 
discontinuous glass transition at the mean-field level has to be
considered, in finite dimension, as a patchwork of local metastable
states. It is important to remark that the length-scale 
$\xi^*$, which diverges at $T_{K}$, plays a
different r{\^o}le than the usual correlation length close to a standard
phase transition. In the latter case, the system appears critical on 
length-scales smaller than the correlation length whereas in the former
case the system is mean-field like for $\xi<\xi^*$. From this
perspective the liquid may
 also be considered as a patchwork of local mean field systems. This
 idea of `blobs'
 is implicitly used in all energy landscape pictures for glassy dynamics
 (see e.g. \cite{Houches,Heuer,Denny,W6}). From the above discussion,
 these pictures 
 certainly cannot describe the dynamics for time scales larger than 
 $\tau(\xi^*)$.\\

 We think that the mosaic picture is more general than originally
 proposed
 by Kirkpatrick, Thirumalai and Wolynes. As an example, consider the
 Kob-Andersen model \cite{KA,BFT} which has a dynamical
 MCT-like transition at the mean-field level (actually on a random
 graph) 
 and no transition but a very rapid increasing of the relaxation
 timescale in finite dimension. The dynamical MCT
 transition is due to the fact that at a certain density $\rho_{MCT}$ 
 an infinite cluster of completely blocked particle appears suddenly.
 As a consequence, for $\rho >\rho_{MCT}$, the configurational space is
 broken up in an exponential number of ergodic components, i.e. there
 is a finite
 configurational entropy \cite{BFT}.
 In finite dimensions, this transition is destroyed: on
 length-scales larger than a (very rapidly increasing) length $\Xi (\rho
 )$ (see \cite{BFT}) one finds with probability of order one 
 some configurations of vacancies that can move cooperatively together.
 Instead, on lengths smaller than $\Xi (\rho)$ the system is completely
 jammed as for mean field systems in the regime $\rho
 >\rho_{MCT}$. Hence, in this case one also finds 
 that the length $\Xi
 (\rho)$ separates a mean-field regime from a non mean-field one.
 Again, the liquid can be considered as a patchwork of local (mean-field
 like)
 metastable states. However, there are two important differences with
 respect to the case discussed in the previous section. First, $\Xi
 (\rho
 )$ is not determined by the competition between energy and entropy.
 Instead 
 what happens is that on length-scale smaller than  $\Xi (\rho
 )$ the configurational space is typically broken up in disconnected
 pieces
 which only start being connected on length-scales of the order of 
 $\Xi (\rho)$ by very rare paths. Second, in the case of the
 Kob-Andersen
 model as well as for many other kinetically constrained models
 \cite{RS}
 the relation between time and length is $\tau \propto \Xi^z$, to be
 contrasted to the exponentially activated one, Eq. (\ref{tauxi}). \\

 We think that for many glass-forming liquids, at least the ones 
 for which the temperature, not the density,
 is the important control parameter inducing the glass transition
 \cite{AlbaSimionesco}, it is
 more likely that the mechanism behind
 the increasing of the relaxation time is the one described in the
 previous section. However,  for other systems like 
 colloids, modeled by hard spheres systems,  
 the mechanism may be different and perhaps similar to the one discussed
 for the Kob-Andersen model \cite{KA,BFT} and present in other
 Kinetically Constrained Models \cite{RS,BFTlong}. In general, one
 expects 
 a mixed scenario where both mechanisms interact to various degrees, 
 with entropic and energetic slowing down entangled (see e. g.
 \cite{Bertin}
 for a toy model).

 \section{Discussion and conclusion}

 The picture proposed in this paper seems to us to be the correct way
 to interpret for real systems the glass transition scenario suggested
 by mean-field models (which should be relevant for physical systems
 where 
 temperature is the relevant parameter governing the glass transition 
 \cite{AlbaSimionesco}). Analytical calculations based on this approach
 \cite{GJPfuture}
 should allow one to obtain quantitatively the length $\xi^*$ for a given
 glass-forming
 liquid, and complement the finite dimensional 
 thermodynamic calculations of \cite{W5,MP,Cardenas}. This would 
 then put on a firmer basis the explanation of the glass transition 
 first proposed by Kirkpatrick, Thirumalai and Wolynes \cite{Wolynes}. 
 In this scenario, the mode-coupling transition temperature 
 $T_{MCT}$ corresponds to
 the appearance of long-lived metastable configurations, 
 which would be stable for infinite range interactions. 
 Above $T_{MCT}$, these (typical equilibrium) 
 states are not even locally stable, much as for spinodal points
 \cite{Wolynes}. 
 Below $T_{MCT}$, these metastable states are mutually accessible, but
 on a small 
 length-scale $\xi$ only a few are relevant, as in the Random Energy
 Model 
 below its glass transition. The length-scale $\xi$ 
 actually plays the r{\^o}le of the temperature in the Random 
 Energy Model; on lengthscales $\xi > \xi^*$ the system is in its liquid
 phase. 
 Our argument above is a mixture of thermodynamics and kinetics:
 small systems are thermodynamically frozen in low energy states,
 whereas large systems are kinetically frozen by
 large activation energies. The relevant relaxation time is that of the
 smallest `liquid' clusters. 
 Mean-field energy landscape pictures that emphasize the importance of
 saddles and valleys/traps -- see,
 e.g.
 \cite{Goldstein,dMnew,B92,KL,Andrea,Houches,Denny,Heuer,Bertin,Review}
 --
 can at best be valid for time scales smaller than this relaxation time,
 i.e. for $\xi < \xi^*$. 
 Interestingly, however, this includes the aging regime (where by 
 definition) the waiting time $t_w$ is smaller than $\tau(\xi^*)$), for
 which 
 one may expect a short 
 time regime where mean-field dynamics \cite{Review} is correct, before
 an intermittent, trap-like
 dynamics takes over. For times greater than $\tau(\xi^*)$, the `cluster
 melting' mechanism discussed here sets in.\\

 Let us emphasize that we have clearly identified a length 
 $\xi^{*}$ beyond which the mean-field picture is not consistent. If
 $\theta>0$ the mean-field picture should still be a good starting point
 to
 describe the physics. The idea of a mosaic state seems to be the
 natural way to adopt the
 mean-field picture in finite dimensions. If instead  $\theta \leq 0$,
 the mean field scenario is completely 
 destroyed, much as for Ising spin systems for example where $\theta >
 0$ only if $d > 1$.  
 Thus, a natural question is the value of $\theta$ for
 glass-forming liquids, but also for all the other systems (pure or
 disordered, classical or quantum) 
 for which a mean field glassy phase has been found \cite{Review}. It
 would therefore be very important to 
 compute $\theta$ in finite dimensions for these models.\\

 A way to try to extract numerically the value of the interface exponent
 $\theta$ for glass-forming 
 liquids and test the mosaic picture would be to simulate the
 thermodynamics of 
 small clusters embedded in a frozen environment chosen to be a typical
 equilibrium 
 configuration of a larger sample. Our prediction
 is that the non ergodicity parameter 
 (height of the plateau of any generic density correlation function) 
 will be large for small $\xi$ and decay rather abruptly to zero beyond
 $\xi^*$, signaling the dominance of configurational entropy
 effects. (See \cite{SKB,KimY} for simulations on glass-forming liquids
 analyzing similar
 effects; note however that these simulations access mainly the
 regime $T>T_{MCT}$ where, within the above framework, $\Upsilon=0$ and
 the relevant dynamical
 lengthscale is the one associated to the MCT transition \cite{BB}
 rather than $\xi^{*}$. 
 Further work would be certainly valuable to conclude whether  $\theta
 >0$ or $\theta < 0$.)
 This crossover scale $\xi^*$ should increase when the temperature is
 decreased, and the relaxation time 
 of the system should grow as an exponential of this cross-over scale
 $\xi^*$. The picture promoted 
 here also suggests that the so-called $\beta$ relaxation regime should
 exhibit non trivial scalings for 
 $T < T_{MCT}$ since it involves motion on all scales $1 \leq \xi \leq
 \xi^*$. 
 The prediction of a non trivial structure of the $\beta$ relaxation in
 the super-cooled regime 
 $T > T_{MCT}$ \cite{Gotze}
 is in fact a spectacular success of Mode Coupling Theory; it was in
 fact recently shown that the nature of the $\beta$
 relaxation in the vicinity of $T_{MCT}$ involves the divergence of a
 dynamical length scale \cite{BB}.
 Finally, dynamical heterogeneities appear very naturally within this
 scenario, first in the MCT regime \cite{BB} and at lower temperatures
 because of the distribution
 of local free-energies within the mosaic.\\

 We should also compare the above configurational entropy scenario with
 other recent proposals. 
 Kivelson, Tarjus and others
 \cite{Tarjus}
 have argued that the slowing down of the dynamics is associated to 
 the proximity of an avoided second order phase transition. 
 Although the local order parameter of this phase transition is not easy
 to define and measure, 
 the existence of a supra-molecular length scale comes from the
 incipient order that is trying to grow below the
 the avoided transition temperature. Although the underlying physics is
 rather different from the 
 above scenario, a great deal of the
 phenomenology is expected to be rather similar. 
 In particular, the system is expected to be frozen (ordered) on short
 scales and `paramagnetic' (liquid) beyond a certain length scale. \\

 Another suggestion is that the slow dynamics is governed by the
 rarefaction 
 of `mobility defects', i.e. simple defects that unlock the dynamics in
 their 
 immediate vicinity. This idea dates back to
 Glarum \cite{Glarum} in the '60s, and was revisited within the
 context of kinetically constrained models by Fredrickson and Andersen
 \cite{FredericksonAndersen}. This picture has been 
 actively promoted in a series of recent papers by 
 Garrahan, Chandler and collaborators \cite{GC1,BG,WGB} (cf. also
 the discussion in the previous section and the review
 \cite{RS}). Although these models lead to dynamics that differ in their
 details, 
 they all posit a complete decoupling between 
 thermodynamics and kinetics, leading to a `non topographic' scenario
 for the glass transition \cite{BG}, 
 in sharp contrast 
 with the potential energy landscape and configurational entropy ideas.
 Thus, these models have nothing to say on the observed Adam-Gibbs
 correlations 
 between configurational entropy and kinetics, which can only be
 accidental from their point of view.
 Roughly speaking the characteristic length scale $\xi^*$ is in 
 this case the typical distance between these rare defects. 
 A small cluster of size $\xi < \xi^*$ typically contains zero
 defects and is kinetically frozen, except for small 
 probability instances where it is fluid. For $\xi > \xi^*$, the
 defects present allow the system to relax on a time scale that behaves
 as 
 a power-law $\xi^{*z}$, rather than an exponential 
 (zero dimensional defects correspond formally to $\psi=0$). 
 This is an important quantitative and qualitative difference with 
 an activated barrier scenario: in order to have very large times
 (as it is the case close to the glass transition) one needs
 very large length-scales. For instance, 
 in order to reach relaxation times $10^{12}$ 
 larger than the microscopic time, $\xi^*$ has to be quite large -- 
 for example, for $z \approx 4$ \cite{WGB}, one obtains a $\xi^*$
 of the order of thousand diameters of particles,
 much larger than in a barrier dominated scenario. 
 Another important difference is the structure of the $\beta$
 relaxation, which
 is expected to be rather trivial in these models. Let us
 note, however, that for other models not characterized by simple
 diffusing defects,
 like the East model, the
 exponent $z$ increases upon lowering $T$ and therefore more modest
 length-scales
 are needed to produce a change of twelve order of magnitude in the
 timescale
 \cite{GC1}. Furthermore the $\beta$ relaxation also turns out to
 be
 non trivial in these models \cite{Ludo}. 
 The fundamental difference with the mosaic state picture
 will then 
 be in the dimensionality of the cooperative regions and,
 correspondingly, on the 
 order of magnitude of the configurational entropy per particle at the
 glass transition -- small
 for diluted defect models and of order $k_B$ for the mosaic state
 \cite{W6}.\\

 We hope that a clear distinction between these pictures will emerge in
 the 
 near future, thanks to the combined efforts 
 of experiments, numerical simulations and theoretical arguments to
 define and measure dynamical lengths in glassy
 systems
 \cite{Ediger,Harrowell,Onuki,Glotzer,Berthier,GC1,WGB,W4,BB,G4inprep}.
 From a theoretical point of view, it is 
 extremely important to substantiate the claims made above 
 by detailed calculations in the context of a
 specific model \cite{GJPfuture}. 
 Models with long-ranged, Kac like interactions look promising in this
 respect \cite{Schmalian,KacFranz}.

 \begin{acknowledgments}
 We thank E. Bertin, L. Berthier, L. Cugliandolo, C. Toninelli, M. Wyart
 for useful discussions. We also thank L. Berthier, J. Kurchan, M.
 M{\'e}zard and D. R. Reichman for comments 
 on the manuscript
 and for many conversations in the past on the physics of glasses,
 including, with M.M., the 
 issue of entropic droplets. 
 \end{acknowledgments}


\begin{thebibliography}{99}
\bibitem{f1} Whether the configurational entropy vanishes at $T_{K}$,
leading to a thermodynamic transition to an amorphous state of matter, or
simply becomes very small,  
is certainly a fascinating theoretical question but mostly irrelevant for 
practical purposes because the experimental region is always at least 
$10 \%$ above $T_K$.
\end{thebibliography}
	\end{document}